\documentstyle[12pt]{article}
\begin{document}
\baselineskip=18pt
\begin{titlepage}
\begin{center}
\large{\bf SOLITON SOLUTIONS FOR THE $N=2$ SUPERSYMMETRIC KDV
EQUATION}
\end{center}
\vspace{2 cm}
\begin{center}
{\bf Sasanka Ghosh}\footnote{email: sasanka@iitg.ernet.in} and
{\bf Debojit Sarma}\footnote{email: debojit@iitg.ernet.in}\\
{\it Department of Physics, Indian Institute of Technology,}\\
{\it North Guwahati, Guwahati 781039, INDIA}\\
\end{center}
\vspace{1 cm}

PACS numbers: 11.30.Pb, 05.45.-a, 05.45.Yv\\

Keywords: $N=2$ superKdV, Hirota method, tau-function.
\vspace{3 cm}
\begin{abstract}
The $N=2$ supersymmetric  KdV equation of Inami and Kanno is 
bilinearized employing the Hirota method and the existence of $N$ 
soliton solutions is demonstrated. The exact form of the solutions are 
explicitly obtained and an interesting symmetry of the equations of 
motion is observed.
\end{abstract}
\end{titlepage}

\newpage
Supersymmetric generalizations of integrable systems have been a field of
considerable interest since the KdV equation was incorporated in a 
supersymmetric framework \cite{Manin,Mathieu}. Apart from being a nonlinear 
equation of 
widespread applicability, it has added significance as equations of the 
KdV type arise in the context of 2D quantum gravity \cite{Zinn}. Moreover 
its supersymmetric extensions are important, since it is believed that a 
fundamental physical theory must possess supersymmetry. Various aspects of 
integrability of supersymmetric equations have been extensively studied 
mainly by extending the Lax formalism to the supersymmetric situation.
The biHamiltonian structures of the superKdV as well as the superKP 
hierarchies have been obtained by suitably formulating the Gelfand Dikii 
method for the supersymmetric systems \cite{SG,Sarma1}. Many of the techniques 
used in the bosonic integrable systems have been extended to the supersymmetric version of the theory. Mention may be made of the Backlund tranformations 
\cite{Kulish}, the Painleve analysis \cite{Bourque}, the Darboux 
transformations \cite{Liu} and the $\tau$-functions \cite{Ibort} vis a vis the 
soliton solutions which have been generalized to the supersymmetric framework. 
However, a powerful and general method of generating soliton solutions in 
bosonic nonlinear differential equations, namely the Hirota bilinear method 
\cite{Hirota} has been applied to supersymmetric extensions of only a few 
nonlinear systems; these have been mainly $N=1$ supersymmetric equations 
\cite{Yung,Carstea}. But the bilinear method for $N=2$ supersymmetric 
equations are yet to be studied.
 
A supersymmetric system of special interest is the $N=2$ KdV equation 
formulated by Inami and Kanno \cite{Inami} from the non-standard superLax 
representation. It was later shown that this system could be obtained from 
the reduction of a $N=2$ superKP hierarchy \cite{Ghosh} arising out of the 
non-standard flow equation. Even though integrability of this system has 
been established 
from the existence of biHamiltonian structures \cite{Sarma1}, its bilinear 
form as well as its soliton solutions are known only when certain conditions 
are imposed on the superfields \cite{Sarma}. In this letter, an attempt is 
made to bilinearize $N=2$ supersymmetric systems, embedded in $N=1$ 
superspace, following Hirota's method and 
to obtain soliton solutions. This will be a 
generalization of the earlier work \cite{Sarma} in the sense that no 
constraints on the superfields have been imposed. As an example, the 
supersymmetric generalization of Hirota bilinear method is considered for 
$N=2$ supersymmetric KdV equation of Inami and Kanno. Specifically, soliton 
solutions of the Inami Kanno equation are obtained by casting it in the 
bilinear form and existence of $N$ soliton solutions are shown to exist 
thereby establishing the integrability of the system.

The Inami Kanno equation \cite{Inami} is given by 
\begin{equation}
\partial_{t}u_{-1}+u^{[6]}_{-1}+3\left(u^{[1]}_{-1}u_{0}\right)^{[2]}
-\frac{1}{2}\left(u^{3}_{-1}\right)^{[2]}=0
\label{1}
\end{equation}
\begin{equation}
\partial_{t}u_{0}+u^{[6]}_{0}-3\left(u_{0}u^{[1]}_{0}\right)^{[2]}-
\frac{3}{2}\left(u_{0}u^{2}_{-1}\right)^{[2]}+3\left(u_{0}u^{[2]}_{-1}
\right)^{[2]}=0
\label{2}
\end{equation}
where $[\;n\;]$ denotes the $n$th derivative with respect to the $D$ operator, 
defined by
\begin{equation}
D=\frac{\partial}{\partial\theta}+\theta\frac{\partial}{\partial x}
\label{3}
\end{equation}
and $u_{-1}$ and $u_{0}$ are superfields of conformal spin $1$ and $3/2$
respectively. It is clear from the form of the equations that they are two
coupled $N=1$ supersymmetric equations which in the limits $u_{-1}=0$ and
$u_{0}=0$ reduce to the $N=1$ KdV equation and the $N=1$ mKdV equation 
\cite{Manin,Mathieu,MBN} respectively. Interestingly, in yet another limit, 
namely $u_{0}=Du_{-1}$ both the equations (\ref{1},\ref{2}) acquire an 
identical form and ultimately reduce to the $N=1$ mKdV equation. 
All these reductions are bilinearizable and $N$ soliton solutions exist for 
each of them \cite{Yung,Carstea,Sarma}. On the other hand, the bosonic limit 
of (\ref{1},\ref{2}) gives rise to a set of coupled equations of the form
\begin{equation}
\partial_{t}u_{-1}^b+\partial^{3}u_{-1}^b-\frac{1}{2}\partial u^{b^3}_{-1}=0
\label{a1}
\end{equation}
and
\begin{equation}
\partial_{t}u_{0}^b+\partial^{3}u_{0}^b-3\partial u_{0}^{b^2}-
\frac{3}{2}\partial\left(u_{0}^b u^{b^2}_{-1}\right)+3\partial
\left[u_{0}^b(\partial u_{-1}^b)\right]=0                                      
\label{a2}
\end{equation}
where the super fields are written in the component forms as
\begin{equation}
u_{-1}=u_{-1}^{b}+\theta u_{-1}^{f}
\label{a}
\end{equation}
and
\begin{equation}
u_{0}=u_{0}^{f}+\theta u_{0}^{b}
\label{b}
\end{equation}                    
The first equation, (\ref{a1}) is nothing but the mKdV equation, whereas the 
second one, (\ref{a2}) is a coupled one and reduces to the KdV equation if 
$u_{-1}^b$ is set to zero. The bosonic equations can be shown to be 
integrable and possess $N$ soliton solutions. 
 
In order to cast the supersymmetric equations (\ref{1},\ref{2}) in the 
bilinear form, we transform the superfields as
\begin{equation}
u_{-1}=2D^{2}\log\frac{\tau_{1}}{\tau_{2}}
\label{4}
\end{equation}
for $u_{-1}$ and
\begin{equation}
u_{0}=2D^{3}\log{\tau_{1}}
\label{5}
\end{equation}
for $u_{0}$. $\tau_{1}$ and $\tau_{2}$  are independent bosonic superfields. 
Notice that in (\ref{5}), $u_0$ depends only the superfield $\tau_1$, 
whereas the other superfield $u_{-1}$ is a function of both $\tau_{1}$ and 
$\tau_{2}$. 

In terms of the supersymmetric Hirota derivative $\bf{S}$ \cite{Yung}, 
defined by
\begin{equation}
{\bf{S}\bf{D}}^{n}_{x}f.g=(D_{\theta_{1}}-D_{\theta_{2}})(\partial_{x_{1}}-
\partial_{x_{2}})^{n}f(x_{1},\theta_{1})g(x_{2},\theta_{2})|_{\begin{array}{c} x_{1}=x_{2}=x \\ 
\theta_{1}=\theta_{2}=\theta\end{array}}
\label{6}
\end{equation}
where $\bf{D}$ is ordinary (or bosonic) Hirota derivative, the $N=2$ super 
KdV equations can be cast in the following bilinear form:
\begin{equation}
({\bf{S}}{\bf{D}}_{t}+{\bf{S}}{\bf{D}}^{3}_{x})(\tau_{1}.\tau_{1})=0
\label{7}
\end{equation}
\begin{equation}
({\bf{S}}{\bf{D}}_{t}+{\bf{S}}{\bf{D}}^{3}_{x})(\tau_{2}.\tau_{2})=0
\label{8}
\end{equation}
\begin{equation}
{\bf{D}}^{2}_{x}(\tau_{1}.\tau_{2})=0
\label{9}
\end{equation}
The bilinear forms in (\ref{7},\ref{8},\ref{9}), indeed, ensure the 
existence of one and two soliton solutions of $N=2$ KdV equations 
(\ref{1},\ref{2}). 

For one soliton solution, we use the the following forms of the 
$\tau$-functions:
\begin{equation}
\tau_{1}=1+\alpha e^{\eta}
\label{10}
\end{equation}
and
\begin{equation}
\tau_{2}=1+\beta e^{\eta}
\label{11}
\end{equation}
$\alpha$ and $\beta$ in (\ref{10},\ref {11}) are nonzero bosonic parameters. 
The parameters 
$\alpha$ and $\beta$ can be determined from the initial value conditions. In 
(\ref{10},\ref {11}), 
\begin{equation}
\eta=kx+\omega t+\zeta \theta
\label{12}
\end{equation}
where $k$ and $\omega$ are the bosonic parameters and $\zeta$ is the 
Grassmann odd parameter. 

Substituting (\ref{10}) and (\ref{11}) back into third equation of the
Hirota form of the superKdV equation (\ref{9}), we find the non-trivial 
solutions exist provided
\begin{equation}
\beta=-\alpha.
\label{13}
\end{equation}
The dispersion relation, however, follows from (\ref{7},\ref{8}) as
\begin{equation}
\omega+k^{3}=0
\label{14}
\end{equation}
which is identical to that for the bosonic KdV or mKdV equations. In 
obtaining the dispersion relation the following property of the superHirota 
operator has been used. 
\begin{equation}
{\bf{S}}{\bf{D}}^{n}\left(e^{\eta_{1}}.e^{\eta_{2}}\right)=\left(k_{1}-k_{2}
\right)^{n}\left[-\left(\zeta_{1}-\zeta_{2}\right)+\theta\left(k_{1}-k_{2}
\right)\right]e^{\eta_{1}+\eta_{2}}
\label{c}
\end{equation}
The fermionic parameter, $\zeta$, however, remains arbitrary at the one 
soliton level. 

The explicit forms of the the soliton solutions for the 
superfields $u_{-1}$ and $u_{0}$ may be found by substituting the $\tau$-
functions in (\ref{4}) and (\ref{5}) respectively. Explicitly
\begin{equation}
u_{-1}=-(2k) {\mbox{cosech}} (\phi + \gamma_0)-\theta (2k\zeta )
{\mbox{cosh}}(\phi + \gamma_0){\mbox{cosech}}^{2}(\phi + \gamma_0)
\label{15}
\end{equation}
and
\begin{equation}
u_{0}=-\frac{1}{2}k\zeta {\mbox{sech}}^{2}\frac{1}{2}
(\phi + \gamma_0)+\theta \frac{k^{2}}{2}{\mbox{sech}}^{2}\frac{1}{2}
(\phi + \gamma_0)
\label{16}
\end{equation}
where we have chosen $\phi = kx-k^3t$ and 
$\alpha=-\beta=e^{\gamma_0}$, $\gamma_0$ being nonzero, real parameter. 
The bosonic and fermionic components 
of the superfields are easily found by comparing (\ref{15}) and (\ref{16}) 
with (\ref{a}) and (\ref{b}). It is observed that the soliton solution of 
$u^b_{-1}$ becomes isomorphic to the mKdV soliton and $u^b_{0}$ to the 
KdV soliton as expected. 

If we choose $\alpha$ to be negative in (\ref{10}), the solutions for 
$u_{-1}$ and $u_{0}$ become
\begin{equation}
u_{-1}=(2k) {\mbox{cosech}} (\phi + \gamma_0)+\theta (2k\zeta )
{\mbox{cosh}}(\phi + \gamma_0){\mbox{cosech}}^{2}(\phi + \gamma_0)
\label{A}
\end{equation}
and
\begin{equation}
u_{0}=\frac{1}{2}k\zeta {\mbox{cosech}}^{2}\frac{1}{2}
(\phi + \gamma_0)-\theta \frac{k^{2}}{2}{\mbox{cosech}}^{2}\frac{1}{2}
(\phi + \gamma_0)
\label{B}
\end{equation}
Interestingly, the solutions (\ref{15},\ref{16}) and (\ref{A},\ref{B}) 
are connected through the transformations 
\begin{equation}
u_{-1}=-u_{-1}
\label{C}
\end{equation}
and 
\begin{equation}
u_{0}=u_{0}-Du_{-1}
\label{D}
\end{equation} 
On the other hand, at the equation level, the  Inami Kanno equations remain 
invariant with respect to the transformations (\ref{C},\ref{D}). 
Evidently, $\alpha\rightarrow -\alpha$ is a symmetry of the the equations 
of motion. At the $\tau$ functions level (\ref{4},\ref{5}), in fact, 
this symmetry is manifested in the tranformation $\tau_1 \rightarrow 
\tau_2$ and vice versa and thus will be observed at the $N$ soliton 
solutions.

The existence of one soliton solution, however, does not
ensure the exact integrability in the Hirota formalism. In order that 
integrability be established, at least three soliton solution must exist.  
The one soliton solution fixes the arbitrary parameters in the trial 
solutions. On the other hand, the interaction term is determined by the two 
soliton solutions, and its form remains unchanged for all higher solitons if 
the system is truly integrable. In fact, if the equations of motion can be 
cast in the bilinear form, one and two soliton solutions can be always 
determined. It is thus necessary to show the existence of at least three 
soliton solution, which checks the consistency of the solution.

The $\tau_1$ for the $N$ soliton of the $N=2$ superKdV may be written as
\begin{equation}
\tau_{1}=\sum_{\mu_{i}=0,1}exp\left(\sum_{i,j=1}^{N}\phi (i,j)\mu_{i}\mu_{j}
+\sum_{i=1}^{N}\mu_{i}(\eta_{i}+\log\alpha_{i})\right)
\;\;\;\;\;(i<j)
\label{24}
\end{equation}
where $A_{ij}=e^{\phi (i,j)}$. For $\tau_{2}$ we replace $\alpha_{i}$ by
$\beta_{i}$ and $A_{ij}$ by $B_{ij}$.
In particular, for the two soliton solution of the superKdV equations 
$\tau$ functions may be written in the form
\begin{equation}
\tau_{1}=1+\alpha_{1}e^{\eta_{1}}+\alpha_{2}e^{\eta_{2}}+\alpha_{1}\alpha_{2} 
A_{12}e^{\eta_{1}+\eta_{2}}
\label{17}
\end{equation}
and
\begin{equation}
\tau_{2}=1+\beta_{1}e^{\eta_{1}}+\beta_{2}e^{\eta_{2}}+\beta_{1}\beta_{2} 
B_{12}e^{\eta_{1}+\eta_{2}}
\label{18}
\end{equation}
where
\begin{equation}
\eta_{1}=k_{1}x+\omega_{1} t+\zeta_{1}\theta
\label{19}
\end{equation}
and
\begin{equation}
\eta_{2}=k_{2}x+\omega_{2} t+\zeta_{2}\theta
\label{20}
\end{equation}
In (\ref{19},\ref{20}) the parameters $k_1$, $k_2$, $\omega_1$ and $\omega_2$ 
are bosonic, while $\zeta_1$ and $\zeta_2$ are fermionic ones, as before.
If we consider $\alpha_1=-\beta_1$, $\alpha_2=-\beta_2$,
\begin{equation}
A_{12}=B_{12}=\frac{(k_{1}-k_{2})^2}{(k_{1}+k_{2})^2}
\label{21}
\end{equation}
and
\begin{equation}
k_{1}\zeta_{2}=k_{2}\zeta_{1}
\label{23}
\end{equation}
the two soliton solutions in (\ref{17},\ref{18}) satisfy the the bilinear 
forms (\ref{7},\ref{8},\ref{9}) leading to the dispersion relations
\begin{equation}
\omega_{1}+k^{3}_{1}=0
\label{d}
\end{equation}
and
\begin{equation}
\omega_{2}+k^{3}_{2}=0
\label{e}
\end{equation}        
The condition (\ref{23}) relating fermionic and bosonic parameters is 
necessary to obtain $A_{12}$ and $B_{12}$, given in (\ref{21}) 
consistently. It will also be seen that the constraint on the 
parameters is crucial in demonstrating the existence of three soliton 
solutions. 

The explicit forms of $\tau_{1}$ and $\tau_{2}$ for the three 
soliton solution are 
\begin{eqnarray}
&&\tau_{1}=1+\alpha_{1}e^{\eta_{1}}+\alpha_{2}e^{\eta_{2}}+\alpha_{3}
e^{\eta_{3}}+\alpha_{1}\alpha_{2} A_{12}e^{\eta_{1}+\eta_{2}}+\alpha_{1}
\alpha_{3} A_{13}e^{\eta_{1}+\eta_{3}}\nonumber\\
&&+\alpha_{2}\alpha_{3} A_{23}e^{\eta_{2}+\eta_{3}}+\alpha_{1}\alpha_{2}
\alpha_{3} A_{12}A_{13}A_{23}e^{\eta_{1}+\eta_{2}+\eta_{3}}
\label{25}
\end{eqnarray}
and
\begin{eqnarray}
&&\tau_{2}=1+\beta_{1}e^{\eta_{1}}+\beta_{2}e^{\eta_{2}}+\beta_{3}e^{\eta_{3}}
+\beta_{1}\beta_{2} B_{12}e^{\eta_{1}+\eta_{2}}+\beta_{1}\beta_{3} 
B_{13}e^{\eta_{1}+\eta_{3}}\nonumber\\
&&+\beta_{2}\beta_{3} B_{23}e^{\eta_{2}+\eta_{3}}+\beta_{1}\beta_{2}\beta_{3} 
B_{12}B_{13}B_{23}e^{\eta_{1}+\eta_{2}+\eta_{3}}
\label{26}
\end{eqnarray}
where
\begin{equation}
\eta_i= k_{ix}x+\omega_i t+\zeta_i \theta \quad(i=1,2,3)
\label{27}
\end{equation}
Notice that the three soliton solution does not contain any new unknown 
parameter; it is expressed in terms of the parameters of the two soliton 
solutions only. These forms of three soliton solution give rise to a set of 
nontrivial relations among the parameters, which determine the consistency of 
the solutions. In the three soliton solution we find the interaction term from 
(\ref{7}) or (\ref{8}), in the same way as for two solitons. Explicitly,
\begin{equation}
A_{ij}=B_{ij}=\frac{(k_{i}-k_{j})^2}{(k_{i}+k_{j})^2} \quad (i,j=1,2,3;\quad
i\neq j)
\label{28}
\end{equation}
This, in turn,  imposes the the following constraint on the fermionic 
parameters:
\begin{equation}
k_{i}\zeta_{j}=k_{j}\zeta_{i} \quad (i,j=1,2,3;\quad i\neq j)                  
\label{29}
\end{equation}
The interaction term $A_{ij}$ can also be found from the third bilinear form 
(\ref{9}) and is identical to (\ref{28}). It is found by direct substitution
that the three soliton solutions (\ref{25}) and (\ref{26}) are consistent with
conditions derived from the one and two soliton solutions. This proves that 
the $N=2$ super KdV equation of Inami and Kanno is also integrable from the 
Hirota point of view.

In conclusion, this is a first attempt to bilinearize $N=2$ supersymmetric
equation and to obtain $N$ soliton solutions. The example considered here
is the $N=2$ KdV equation of Inami and Kanno which has two interesting $N=1$
limits. If $u_{-1}$ is set to zero, it reduces to $N=1$ KdV equation, whereas
by choosing the limit $u_{0}=0$ or $u_{0}=Du_{-1}$, it becomes $N=1$ mKdV
equation. The fermionic parameter $\zeta$ plays a trivial role at the one 
soliton solution level, but at the higher soliton level it is intimately 
related with the bosonic parameters and eventually the conditions become
essential to show the existence of $N$ soliton solutions vis a vis 
the integrability of the system.  
\vspace{1 cm}

{\it SG would like to thank DST, Govt. of India for financial support under 
the project no. 100/(IFD)/2066/2000-2002.}

\end{document}